\documentclass[runningheads]{llncs}
\usepackage{graphicx}
\usepackage{comment}
\usepackage{amsmath,amssymb} 
\usepackage{color}

\usepackage{subfigure}
\usepackage{tabularx}
\usepackage{booktabs}
\usepackage{multirow}
\usepackage{pifont}
\newcommand{\cmark}{\ding{51}}
\newcommand{\xmark}{\ding{55}}
\usepackage[colorlinks,urlcolor=blue,driverfallback=dvipdfm]{hyperref}
\graphicspath{{figures/}}

\begin{document}
\pagestyle{headings}
\mainmatter
\def\ECCVSubNumber{6628}  

\title{Cross-Attention in Coupled Unmixing Nets for Unsupervised Hyperspectral Super-Resolution} 

\titlerunning{Coupled Unmixing Nets with Cross-Attention}
%
\author{Jing Yao\inst{1,2,4}\orcidID{0000-0003-1301-9758} \and
Danfeng Hong\inst{2}\thanks{Corresponding author}\orcidID{0000-0002-3212-9584} \and \\
Jocelyn Chanussot\inst{3}\orcidID{0000-0003-4817-2875} \and
Deyu Meng \inst{1,5}\orcidID{0000-0002-1294-8283} \and \\
Xiaoxiang Zhu \inst{2,4}\orcidID{0000-0001-5530-3613}\and
Zongben Xu \inst{1}}
\authorrunning{J. Yao et al.}
%
\institute{School of Mathematics and Statistics, Xi'an Jiaotong University, China
\email{jasonyao@stu.xjtu.edu.cn; \{dymeng, zbxu\}@mail.xjtu.edu.cn}\\
\and
Remote Sensing Technology Institute, German Aerospace Center, Germany\\
\email{\{danfeng.hong, xiaoxiang.zhu\}@dlr.de}
\and
Univ. Grenoble Alpes, INRIA, CNRS, Grenoble INP, LJK, France\\
\email{jocelyn.chanussot@grenoble-inp.fr}
\and
Technical University of Munich, Germany\\
\and
Macau University of Science and Technology, China}
\maketitle

\begin{abstract}
The recent advancement of deep learning techniques has made great progress on hyperspectral image super-resolution (HSI-SR). Yet the development of unsupervised deep networks remains challenging for this task. To this end, we propose a novel coupled unmixing network with a cross-attention mechanism, CUCaNet for short, to enhance the spatial resolution of HSI by means of higher-spatial-resolution multispectral image (MSI). Inspired by coupled spectral unmixing, a two-stream convolutional autoencoder framework is taken as backbone to jointly decompose MS and HS data into a spectrally meaningful basis and corresponding coefficients. CUCaNet is capable of adaptively learning spectral and spatial response functions from HS-MS correspondences by enforcing reasonable consistency assumptions on the networks. Moreover, a cross-attention module is devised to yield more effective spatial-spectral information transfer in networks. Extensive experiments are conducted on three widely-used HS-MS datasets in comparison with state-of-the-art HSI-SR models, demonstrating the superiority of the CUCaNet in the HSI-SR application. Furthermore, the codes and datasets are made available at: \url{https://github.com/danfenghong/ECCV2020_CUCaNet}.
\end{abstract}

\section{Introduction}
Recent advances in hyperspectral (HS) imaging technology have enabled the availability of enormous HS images (HSIs) with a densely sampled spectrum \cite{rasti2020feature}. Benefited from the abundant spectral information contained in those hundreds of bands measurement, HSI features great promise in delivering faithful representation of real-world materials and objects. Thus the pursuit of effective and efficient processing of HS data has long been recognized as a prominent topic in the field of computer vision \cite{fu2018joint,gao2020spectral}.

Though physically, the insufficient spatial resolution of HS instruments, combined with an inherently intimate mixing effect, severely hampers the abilities of HSI in various real applications \cite{akhtar2014sparse,yao2019nonconvex}. Fortunately, the multispectral (MS) imaging systems (e.g., RGB cameras, spaceborne MS sensors) are capable of providing complementary products, which preserve much finer spatial information at the cost of reduced spectral resolution \cite{hong2015novel}. Accordingly, the research on enhancing the spatial resolution (henceforth, resolution refers to the spatial resolution) of an observable low-resolution HSI (LrHSI) by merging a high-resolution MSI (HrMSI) under the same scene, which is referred to hyperspectral image super-resolution (HSI-SR), has been gaining considerable attention \cite{hong2019cospace,hong2019learnable}.

The last decade has witnessed a dominant development of optimization-based methods, from either deterministic or stochastic perspectives, to tackle the HSI-SR issue \cite{yokoya2017hyperspectral}. To mitigate the severe ill-posedness of such an inverse problem, the majority of prevailing methods put their focus on exploiting various hand-crafted priors to characterize spatial and spectral information underlying the desired solution. Moreover, the dependency on the knowledge of relevant sensor characteristics, such as spectral response function (SRF) and point spread function (PSF), inevitably compromises their transparency and practicability.

More recently, a growing interest has been paid to leverage the tool of deep learning (DL) by exploiting its merit on low-level vision applications. Among them, the best result is achieved by investigators who resort to performing HSI-SR progressively in a supervised fashion \cite{xie2019multispectral}. However, the demand for sufficient training image pairs acquired with different sensors inevitably makes their practicability limited. On the other hand, though being rarely studied, the existing unsupervised works rely on either complicated multi-stage alternating optimization \cite{qu2018unsupervised}, or an external camera spectral response (CSR) dataset in the context of RGB image guidance \cite{fu2019hyperspectral}, the latter of which also losses generality in confronting other kinds of data with higher spectral resolution than RGB one.

To address the aforementioned challenges, we propose a novel coupled unmixing network with cross-attention (CUCaNet) for unsupervised HSI-SR. The contributions of this paper are briefly summarized as follows:
\begin{enumerate}
    \item We propose a novel unsupervised HSI-SR model, called CUCaNet, which is built on a coupled convolutional autoencoder network. CUCaNet models the physically mixing properties in HS imaging into the networks to transfer the spatial information of MSI to HSI and preserve the high spectral resolution itself simultaneously in a coupled fashion. 
    \item We devise an effective cross-attention module to extract and transfer significant spectral (or spatial) information from HSI (or MSI) to another branch, yielding more sufficient spatial-spectral information blending.
    \item Beyond previous coupled HSI-SR models, the proposed CUCaNet is capable of adaptively learning PSFs and SRFs across MS-HS sensors with a high ability to generalize. To find the local optimum of the network more effectively, we shrink the solution space by designing a closed-loop consistency regularization in networks, acting on both spatial and spectral domains. 
\end{enumerate}

\section{Related Work}
Pioneer researches have emerged naturally by adapting the similar but extensively studied pansharpening techniques to HSI-SR \cite{vivone2014critical,loncan2015hyperspectral}, which usually fail to well capture the global continuity in the spectral profiles and thus brings unignorable performance degradation, leaving much room to be desired.

\subsection{Conventional Methods}
Apace with the advances in statistically modeling and machine learning, recent optimization-based methods has lifted the HSI-SR ratio evidently. According to a subspace assumption, Bayesian approach was first introduced by Eismann \textit{et al.} utilizing a stochastic mixing model \cite{eismann2004resolution}, and developed through subsequent researches by exploiting more inherent characteristics \cite{simoes2014convex,wei2015fast}. Another class of methods that have been actively investigated stems from the idea of spectral unmixing \cite{hong2019augmented}, which takes the intimate mixing effect into consideration. Yokoya \textit{et al.} brought up coupled non-negative matrix factorization (CNMF) \cite{yokoya2011coupled} to estimate the spectral signature of the underlying materials and corresponding coefficients alternately. On basis of CNMF, Kawakami \textit{et al.} \cite{bieniarz2011hyperspectral} employed sparse regularization and  an effective projected gradient solver was devised by Lanaras \textit{et al.} \cite{lanaras2015hyperspectral}. Besides, \cite{akhtar2014sparse,dong2016hyperspectral} adopted dictionary learning and sparse coding techniques in this context. Various kinds of tensor factorization strategies are also studied, such as Tucker decomposition adopted by Dian \textit{et al.} \cite{dian2019learning} and Li \textit{et al.} \cite{li2018fusing} to model non-local and coupled structure information, respectively.

\subsection{DL-Based Methods}
To avoid tedious hand-crafted priors modeling in conventional methods, DL-based methods have attracted increasing interest these years. In the class of supervised methods, Dian \textit{et al.} \cite{dian2018deep} employed CNN with prior training to finely tune the result acquired by solving a conventional optimization problem, while Xie \textit{et al.} \cite{xie2019multispectral} introduced a deep unfolding network based on a novel HSI degradation model. Unsupervised methods are more rarely studied. Qu \textit{et al.} \cite{qu2018unsupervised} developed an unsupervised HSI-SR net with Dirichlet distribution-induced layer embedded, which results in a multi-stage alternating optimization. Under the guidance of RGB image and an external CSR database, Fu \textit{et al.} \cite{fu2019hyperspectral} designed an unified CNN framework with a particular CSR optimization layer. Albeit demonstrated to be comparatively effective, these methods require either large training data for supervision or the knowledge of PSFs or SRFs, which are both unrealistic in real HSI-SR scenario. Very recently, Zheng \textit{et al.} \cite{zhen2020coupled} proposed a coupled CNN by adaptively learning the two functions of PSFs and SRFs for unsupervised HSI-SR. However, due to the lack of effective regularizations or constraints, the two to-be-estimated functions inevitably introduce more freedoms, limiting the performance to be further improved.

\begin{figure*}[!t]
    \centering
	\subfigure{
		\includegraphics[width=1\textwidth]{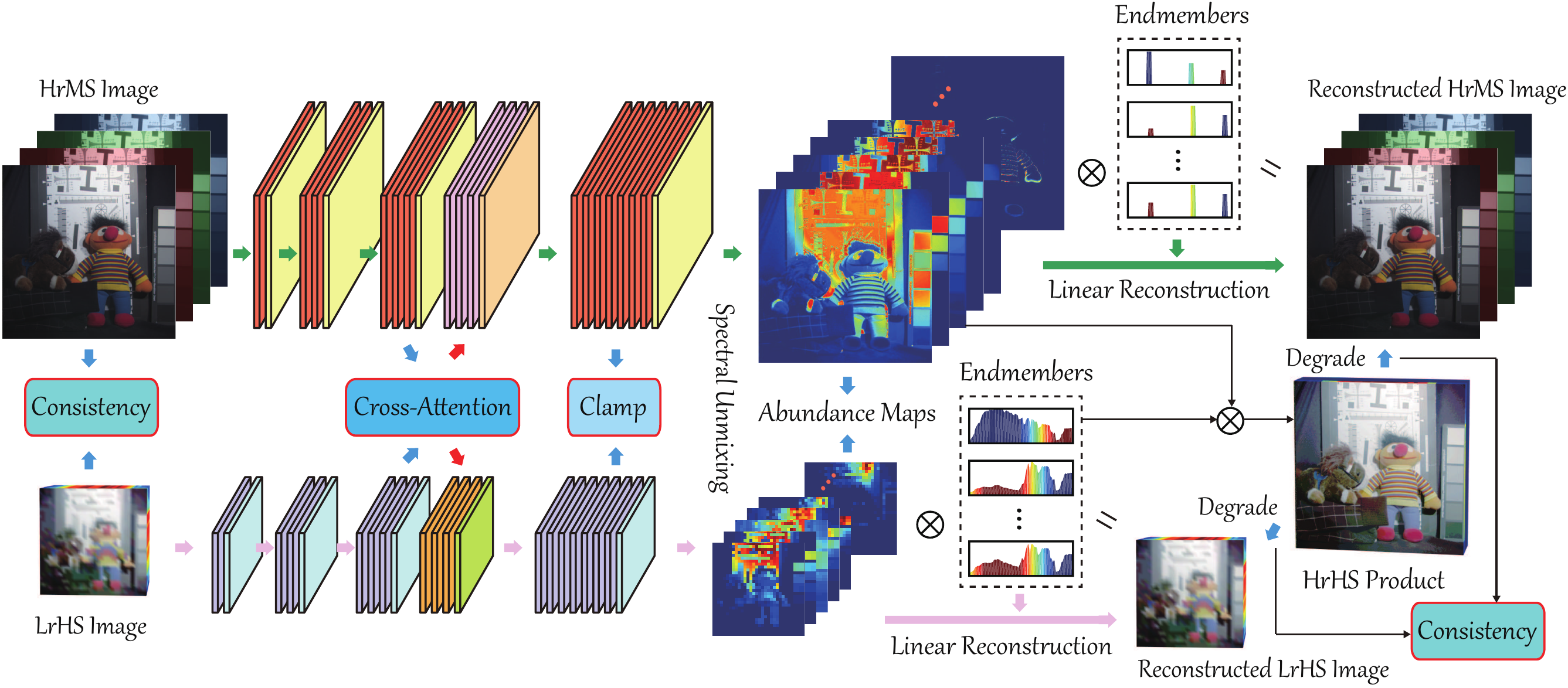}
	}
    \caption{An illustration of the proposed end-to-end CUCaNet inspired by spectral unmixing techniques, which mainly consists of two important modules: cross-attention and spatial-spectral consistency. }
    \label{fig:overview}
\end{figure*}

\section{Coupled Unmixing Nets with Cross-Attention}
In this section, we present the proposed coupled unmixing networks with a cross-attention module implanted, which is called CUCaNet for short. For mathematical brevity, we resort to a 2D representation of the 3D image cube, that is, the spectrum of each pixel is stacked row-by-row. 

\subsection{Method Overview}
CUCaNet builds on a two-stream convolutional autoencoder backbone, which aims at decomposing MS and HS data into a spectrally meaningful basis and corresponding coefficients jointly. Inspired by CNMF, the fused HrHSI is obtained by feeding the decoder of the HSI branch with the encoded maps of the MSI branch. Two additional convolution layers are incorporated to simulate the spatial and spectral downsampling processes across MS-HS sensors. To guarantee that CUCaNet can converge to a faithful product through an unsupervised training, reasonable consistency, and necessary unmixing constraints, are integrated smoothly without imposing evident redundancy. Moreover, we introduced the cross-attention attention mechanism into the HSI-SR for the first time.

\begin{figure*}[!t]
    \centering
	\subfigure{
		\includegraphics[width=0.95\textwidth]{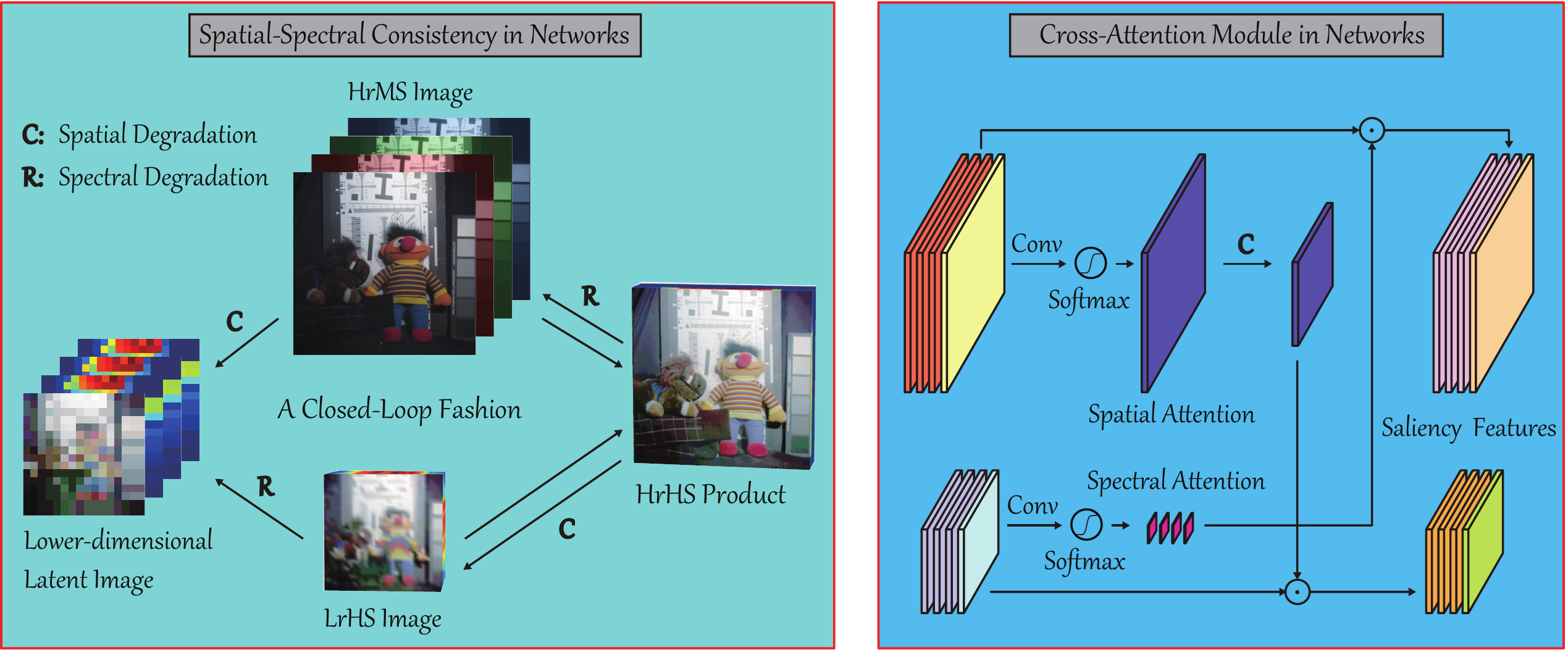}
	}
    \caption{Detail unfolding for two modules in networks: spatial-spectral consistency (left) and cross-attention (right).}
    \label{fig:modules}
\end{figure*}

\subsection{Problem Formulation}\label{SC:problem}
Given the LrHSI $\mathbf{X}\in\mathbb{R}^{hw\times L}$, and the HrMSI $\mathbf{Y}\in\mathbb{R}^{HW\times l}$, the goal of HSI-SR is to recover the latent HrHSI $\mathbf{Z}\in\mathbb{R}^{HW\times L}$, where $(h,w,s)$ are the reduced height, width, and number of spectral bands, respectively, and $(H,W,S)$ are corresponding upsampled version. Based on the linear mixing model that well explains the phenomenon of \textit{mixed pixels} involved in $\mathbf{Z}$, we then have the following NMF-based representation, 
\begin{equation}\label{Eq:SU}
    \mathbf{Z}=\mathbf{S}\mathbf{A},
\end{equation}
where $\mathbf{A}\in\mathbb{R}^{K\times L}$ and $\mathbf{S}\in\mathbb{R}^{HW\times K}$ are a collection of spectral signatures of pure materials (or say, endmembers) and their fractional coefficients (or say, abundances), respectively. 

On the other hand, the degradation processes in the spatial ($\mathbf{X}$) and the spectral ($\mathbf{Y}$) observations can be modeled as
\begin{align}
    & \mathbf{X}\approx\mathbf{C}\mathbf{Z}=\mathbf{C}\mathbf{S}\mathbf{A}=\tilde{\mathbf{S}}\mathbf{A},\label{Eq:X}\\
    & \mathbf{Y}\approx\mathbf{Z}\mathbf{R}=\mathbf{S}\mathbf{A}\mathbf{R}=\mathbf{S}\tilde{\mathbf{A}},\label{Eq:Y}
\end{align}
where $\mathbf{C}\in\mathbb{R}^{hw\times HW}$ and $\mathbf{R}\in\mathbb{R}^{L\times l}$ represent the PSF and SRF from the HrHSI to the HrMSI and the LrHSI, respectively. Since $\mathbf{C}$ and $\mathbf{R}$ are non-negative and normalized, $\tilde{\mathbf{S}}$ and $\tilde{\mathbf{A}}$ can be regarded as spatially downsampled abundances and spectrally downsampled endmembers, respectively. Therefore, an intuitive solution is to unmix $\mathbf{X}$ and $\mathbf{Y}$ based on Eq. (\ref{Eq:X}) and Eq. (\ref{Eq:Y}) alternately, which is coupled with the prior knowledge of $\mathbf{C}$ and $\mathbf{R}$. Such a principle has been exploited in various optimization formulations, obtaining state-of-the-art fusion performance by linear approximation with converged $\mathbf{S}$ and $\mathbf{A}$.

\textbf{Constraints.}~Still, the issued HSI-SR problem involves the inversions from $\mathbf{X}$ and $\mathbf{Y}$ to $\mathbf{S}$ and $\mathbf{A}$, which are highly ill-posed. To narrow the solution space, several physically meaningful constraints are commonly adopted, they are the abundance sum-to-one constraint (ASC), the abundance non-negative constraint (ANC), and non-negative constraint on endmembers, i.e.,
\begin{equation}\label{Eq:constraints}
    \mathbf{S}\mathbf{1}_K=\mathbf{1}_{HW},~\mathbf{S}\succeq 0,~\mathbf{A}\succeq 0,
\end{equation}
where $\succeq$ marks element-wise inequality, and $\mathbf{1}_p$ represents $p$-length all-one vector. It is worth mentioning that the combination of ASC and ANC would promote the sparsity of abundances, which well characterizes the rule that the endmembers are sparsely contributing to the spectrum in each pixel.

Yet in practice, the prior knowledge of PSFs and SRFs for numerous kinds of imaging systems is hardly available. This restriction motivates us to extend the current coupled unmixing model to a fully end-to-end framework, which is only in need of LrHSI and HrMSI. To estimate $\mathbf{C}$ and $\mathbf{R}$ in an unsupervised manner, we introduce the following consistency constraint,
\begin{equation}\label{Eq:W}
    \mathbf{U}=\mathbf{X}\mathbf{R}=\mathbf{C}\mathbf{Y},
\end{equation}
where $\mathbf{U}\in\mathbb{R}^{hw\times l}$ denotes the latent LrMSI.

\subsection{Network Architecture}
Inspired by the recent success of deep networks on visual processing tasks, we would like to first perform coupled spectral unmixing by the established two-stream convolutional autoencoder for the two-modal inputs, i.e., we consider two deep subnetworks, with $f(\mathbf{X})=f_{de}(f_{en}(\mathbf{X};\mathbf{W}_{f,en});\mathbf{W}_{f,de})$ to self-express the LrHSI, $g(\mathbf{Y})=g_{de}(g_{en}(\mathbf{Y};\mathbf{W}_{g,en});\mathbf{W}_{g,de})$ for the HrMSI, and the fused result can be obtained by $\hat{\mathbf{Z}}=f_{de}(g_{en}(\mathbf{Y};\mathbf{W}_{g,en});\mathbf{W}_{f,de})$, herein $\mathbf{W}$ collects the weights of corresponding subpart.

As shown in Fig. \ref{fig:overview}, both encoders $f_{en}$ and $g_{en}$ are constructed by cascading ``Convolution+LReLU'' blocks $f_l$ with an additional $1\times1$ convolution layer. We set the sizes of convolutional kernels in $f_{en}$ all as $1\times1$ while those in $g_{en}$ are with larger but descending scales of the receptive field. The idea behind this setting is to consider the low fidelity of spatial information in LrHSI and simultaneously map the cross-channel and spatial correlations underlying HrMSI. Furthermore, to ensure that the encoded maps are able to possess the properties of abundances, an additional activation layer using the clamp function in the range of $[0,1]$ is concatenated after each encoder. As for the structure of decoders $f_{de}$ and $g_{de}$, we simply adopt a $1\times1$ convolution layer without any nonlinear activation, making the weights $\mathbf{W}_{f,de}$ and $\mathbf{W}_{g,de}$ interpretable as the endmembers $\mathbf{A}$ and $\tilde{\mathbf{A}}$ according to Eq. (\ref{Eq:X}) and Eq. (\ref{Eq:Y}). By backward gradient descent-based optimization, our backbone network can not only avoid the need for good initialization for conventional unmixing algorithms but also enjoy the amelioration brought by its capability of local perception and nonlinear processing. 

\textbf{Cross-Attention.}~To further exploit the advantageous information from the two modalities, we devise an effective cross-attention module to enrich the features across modalities. As shown in Fig. \ref{fig:modules}, the cross-attention module is employed on high-level features within the encoder part, with three steps to follow. First, we compute the spatial and spectral attention from the branch of LrHSI and HrMSI, since they can provide with more faithful spatial and spectral guidance. Next, we multiply the original features with the attention maps from another branch to transfer the significant information. Lastly, we concatenate the original features with the above cross-multiplications in each branch, to construct the input of next layer in the form of such preserved and refined representation. 

Formally, the output features $\mathbf{F}_l\in\mathbb{R}^{h\times w}$ of the $l$-th layer in the encoder part, take $f_{en}$ for example, are formulated as
\begin{equation}
    \mathbf{F}_l=f_l(\mathbf{F}_{l-1})=f_l(f_{l-1}(\cdots f_1(\mathbf{X})\cdots)),
\end{equation}
which is similar for obtaining $\mathbf{G}_l\in\mathbb{R}^{H\times W}$ from $g_{en}$. To gather the spatial and spectral significant information, we adopt global and local convolution to generate channel-wise and spatial statistics respectively as
\begin{equation}
    o_c=\mathbf{u}_c\odot\mathbf{F}_l^{(c)},~\mathbf{S}=\sum_{c=1}^C\mathbf{v}^{(c)}\odot\mathbf{G}_l^{(c)},
\end{equation}
where $\mathbf{u}=[\mathbf{u}_1,\cdots,\mathbf{u}_C]$ is a set of convolution filters with size $h\times w$, $\mathbf{v}^{(c)}$ is the $c$-th channel of a 3D convolution filter with spatial size as $p\times p$. Then we apply a softmax layer to the above statistics to get the attention maps $\delta(\mathbf{o})\in\mathbb{R}^{C}$, and $\delta(\mathbf{S})\in\mathbb{R}^{H\times W}$, where $\delta(\cdot)$ denotes the softmax activation function. The original features are finally fused into the input of next layer as $concat(\mathbf{F}_l;\mathbf{F}_l\odot\delta(\mathbf{S}))$, and $concat(\mathbf{G}_l;\mathbf{G}_l\odot\delta(\mathbf{o}))$, where $concat(\cdot)$ denotes the concatenation, and $\odot$ denotes the point-wise multiplication.

\textbf{Spatial-Spectral Consistency.}~An essential part that tends to be ignored is related to the coupled factors caused by PSFs and SRFs. Previous researches typically assume an ideal average spatial downsampling and the prior knowledge of SRFs, which rarely exist in reality. Unlike them, we introduce a spatial-spectral consistency module into networks in order to better simulate the to-be-estimated PSF and SRF, which is performed by simple yet effective convolution layers.

We can rewrite the spectral resampling from the HS sensor to the MS sensor by revisiting the left part of Eq. (\ref{Eq:Y}) more accurately as follows. Given the spectrum of $i$-th pixel in HrHSI $\mathbf{z}_{i}$, for the $j$-th channel in corresponding LrHSI, the radiance $y_{i,j}$ is defined as
\begin{equation}
    y_{i,j}=\int_{\phi}\mathbf{z}_i(\mu)\mathbf{r}_j(\mu)d\mu/N_r,
\end{equation}
where $\phi$ denotes the support set that the wavelength $\mu$ belongs to, $N_r$ denotes the normalization constant $\int\mathbf{r}_j(\mu)d\mu$. We directly replace $\mathbf{r}_j$ with a set of $L$ $1\times1$ convolution kernels with the weights being collected in $\mathbf{w}_j$. Therefore, the SRF layer $f_r$ can be well defined as follows,
\begin{equation}
    y_{i,j}=f_r(\mathbf{z}_i;\mathbf{w}_j)=\sum_\phi\mathbf{z}_i(\mu)\mathbf{w}_j(\mu)/N_w,
\end{equation}
where $N_w$ corresponds to an additional normalization with $\sum_\phi\mathbf{w}_j$. The PSF layer for spatial downsampling is more straightforward. Note that PSF generally indicates that each pixel in LrHSI is produced by combining neighboring pixels in HrHSI with unknown weights in a disjoint manner \cite{wang2017effect}. To simulate this process, we propose $f_s$ by the means of a channel-wise convolution layer with kernel size and stride both same as the scaling ratio.

To sum up, multiple consistency constraints derived from the statements in Section \ref{SC:problem}, either spectrally or spatially, can be defined in our networks as
\begin{equation}\label{Eq:SSC}
    \hat{\mathbf{Y}}=f_r(\hat{\mathbf{Z}}),~\hat{\mathbf{X}}=f_s(\hat{\mathbf{Z}}),~f_s(\mathbf{Y})=f_r(\mathbf{X}),
\end{equation}
which enables the whole networks to be trained within a closed loop.

\subsection{Network Training}
\textbf{Loss Function.}~As shown in Fig. \ref{fig:overview}, our CUCaNet mainly consists of two autoencoders for hyperspectral and multispectral data, respectively, thus leading to the following reconstruction loss:
\begin{equation}
    \mathcal{L}_\text{R}=\lVert f(\mathbf{X})-\mathbf{X}\rVert_1+\lVert g(\mathbf{Y})-\mathbf{Y}\rVert_1,
\end{equation}
in which the $\ell_1$-norm is selected as the loss criterion for its perceptually satisfying performance in the low-level image processing tasks \cite{zhao2016loss}.

The important physically meaningful constraints in spectral unmixing are considered, building on Eq. (\ref{Eq:constraints}), we then derive the second ASC loss as
\begin{equation}
    \mathcal{L}_\text{ASC}=\lVert\mathbf{1}_{hw}-f_{en}(\mathbf{X})\mathbf{1}_K\rVert_1+\lVert\mathbf{1}_{HW}-g_{en}(\mathbf{Y})\mathbf{1}_K\rVert_1,
\end{equation}
and the ANC is reflected through the activation layer used behind the encoders.

To promote the sparsity of abundances of both stream, we adopt the Kullback-Leibler (KL) divergence-based sparsity loss term by penalizing the discrepancies between them and a tiny scalar $\epsilon$,
\begin{equation}
    \mathcal{L}_\text{S}=\sum_n\text{KL}(\epsilon||(f_{en}(\mathbf{X}))_n)+\sum_m\text{KL}(\epsilon||(g_{en}(\mathbf{Y}))_m),
\end{equation}
where $\text{KL}(\rho||\hat{\rho})=\rho\log\frac{\rho}{\hat{\rho}}+(1-\rho)\log\frac{1-\rho}{1-\hat{\rho}}$ is the standard KL divergence \cite{ng2011sparse}.

Last but not least, we adopt the $\ell_1$-norm to define the spatial-spectral consistency loss based on Eq. (\ref{Eq:SSC}) as follows,
\begin{equation}
    \mathcal{L}_\text{C}=\lVert f_s(\mathbf{Y})-f_r(\mathbf{X})\rVert_1+\lVert\hat{\mathbf{X}}-\mathbf{X}\rVert_1+\lVert\hat{\mathbf{Y}}-\mathbf{Y}\rVert_1.
\end{equation}

By integrating all the above-mentioned loss terms, the final objective function for the training of CUCaNet is given by
\begin{equation}
    \mathcal{L}=\mathcal{L}_\text{R}+\alpha\mathcal{L}_\text{ASC}+\beta\mathcal{L}_\text{S}+\gamma\mathcal{L}_\text{C},
\end{equation}
where we use $(\alpha, \beta, \gamma)$ to trade-off the effects of different constituents.

\textbf{Implementation Details.}~Our network is implemented on PyTorch framework. We choose Adam optimizer under default parameters setting for training with the training batch parameterized by 1 \cite{kingma2014adam}. The learning rate is initialized with 0.005 and a linear decay from 2000 to 10000 epochs drop-step schedule is applied \cite{loshchilov2018decoupled}. We adopt Kaiming's initialization for the convolutional layers \cite{he2015delving}. The hyperparameters are determined using a grid search on the validation set and training will be early stopped before validation loss fails to decrease.

\section{Experimental Results}
In this section, we first review the HSI-MSI datasets and setup adopted in our experiments. Then, we provide an ablation study to verify the effectiveness of the proposed modules. Extensive comparisons with the state-of-the-art methods on indoor and remotely sensed images are reported at last.

\textbf{Dataset and Experimental Setting.}~Three widely used HSI-MSI datasets are investigated in this section, including CAVE dataset \cite{yasuma2010generalized}\footnote[1]{http://www.cs.columbia.edu/CAVE/databases/multispectral}, Pavia University dataset, and Chikusei dataset \cite{yokoya2017hyperspectral}\footnote[2]{http://naotoyokoya.com/Download.html}. The CAVE dataset captures 32 different indoor scenes. Each image consists of 512$\times$512 pixels with 31 spectral bands uniformly measured in the wavelength ranging from 400nm to 700nm. In our experiments, 16 scenes are randomly selected to report performance. The Pavia dataset was acquired by ROSIS airborne sensor over the University of Pavia, Italy, in 2003. The original HSI comprises 610$\times$340 pixels and 115 spectral bands. We use the top-left corner of the HSI with 336$\times$336 pixels and 103 bands (after removing 12 noisy bands), covering the spectral range from 430nm to 838nm. The Chikusei dataset was taken by a Visible and Near-Infrared (VNIR) imaging sensor over Chikusei, Japan, in 2014. The original HSI consists of 2,517$\times$2,335 pixels and 128 bands with a spectral range of 363nm to 1,018nm. We crop 6 non-overlapped parts with size of 576$\times$448 pixels from the bottom part for test.

Considering the diversity of MS sensors in generating the HrMS images, we employ the SRFs of Nikon D700 camera\cite{qu2018unsupervised} and Landsat-8 spaceborne MS sensor\cite{barsi2014spectral}\footnote[3]{http://landsat.gsfc.nasa.gov/?p=5779} for the CAVE dataset and two remotely sensed datasets\footnote[4]{We select the spectral radiance responses of blue-green-red(BGR) bands and BGR-NIR bands for the experiments on Pavia and Chikusei datasets, respectively.}, respectively. We adopt the Gaussian filter to obtain the LrHS images, by constructing the filter with the width same as SR ratio and 0.5 valued deviations. The SR ratios are set as 16 for the Pavia University dataset  and 32 for the other two datasets.

\textbf{Evaluation Metrics.}~We use the following five complementary and widely-used picture quality indices (PQIs) for the quantitative HSI-SR assessment, including peak signal-to-noise ratio (PSNR), spectral angle mapper (SAM) \cite{kruse1993spectral}, erreur relative globale adimensionnellede synth\`{e}se (ERGAS) \cite{wald2000quality}, structure similarity (SSIM) \cite{wang2004image}, and universal image quality index (UIQI) \cite{wang2002universal}. SAM reflects the spectral similarity by calculating the average angle between two vectors of the estimated and reference spectra at each pixel. PSNR, ERGAS, and SSIM are mean square error (MSE)-based band-wise PQIs indicating spatial fidelity, global quality, and perceptual consistency, respectively. UIQI is also band-wisely used to measure complex distortions among monochromatic images.

\begin{table}[!t]
    \centering
    \caption{Ablation study on the Pavia University dataset by our CUCaNet with different modules and a baseline CNMF. The best results are shown in bold.}
    \resizebox{0.7\textwidth}{!}{
    \begin{tabular}{c|ccc|ccccc}
        \toprule[1.5pt]
        \multirow{2}{*}{Method} & \multicolumn{3}{c|}{Module} & \multicolumn{5}{c}{Metric} \\
        \cline{2-9}
                & ~Clamp & ~SSC~  & ~CA~   &~PSNR  &~SAM~ & ERGAS &~SSIM~ & UQI \\
        \hline\hline
        CNMF    & -      & -      & -      & 32.73 & 7.05 & 1.18  & 0.830 & 0.973 \\
        \hline
        CUCaNet & \xmark & \xmark & \xmark & 34.25 & 6.58 & 1.01  & 0.862 & 0.975 \\
        CUCaNet & \cmark & \xmark & \xmark & 35.67 & 5.51 & 0.92  & 0.897 & 0.981 \\
        CUCaNet & \cmark & \cmark & \xmark & 36.55 & 4.76 & 0.85  & 0.904 & \textbf{0.991} \\
        CUCaNet & \cmark & \xmark & \cmark & 36.49 & 4.63 & 0.86  & 0.902 & 0.989 \\
        \hline
        CUCaNet & \cmark & \cmark & \cmark & \textbf{37.22} & \textbf{4.43} & \textbf{0.82}  & \textbf{0.914} & \textbf{0.991} \\
        \bottomrule[1.5pt]
    \end{tabular}
    }
    \label{table:Ablation}
\end{table}

\subsection{Ablation Study}
Our CUCaNet consists of a baseline network -- coupled convolutional autoencoder networks -- and two newly-proposed modules, i.e., the spatial-spectral consistency module (SSC) and the cross-attention module (CA). To investigate the performance gain of different components in networks, we perform ablation analysis on the Pavia University dataset. We also study the effect of replacing clamp function with conventional softmax activation function at the end of each encoder. Table \ref{table:Ablation} details the quantitative results, in which CNMF is adopted as the baseline method. 

As shown in Table \ref{table:Ablation}, single CUCaNet can outperform CNMF in all metrics owing to its benefit from employing deep networks. We find that the performance is further improved remarkably by the use of clamp function. Meanwhile, single SSC module performs better than single CA module except in SAM, which means that CA module tend to favor spectral consistency. By jointly employing the two modules, the proposed CUCaNet achieves the best results in HSI-SR tasks, demonstrating the effectiveness of our whole network architecture.

\subsection{Comparative Experiments}
\textbf{Compared Methods.}~Here, we make comprehensive comparison with the following eleven state-of-the-art (SOTA) methods in HSI-RS tasks: pioneer work, GSA \cite{aiazzi2007improving}\footnotemark, NMF-based approaches, CNMF \cite{yokoya2011coupled}\footnotemark[\value{footnote}]\footnotetext{http://naotoyokoya.com/Download.html} and CSU \cite{lanaras2015hyperspectral}\footnote{https://github.com/lanha/SupResPALM}, Bayesian-based approaches, FUSE \cite{wei2015fast}\footnote{https://github.com/qw245/BlindFuse} and HySure \cite{simoes2014convex}\footnote{https://github.com/alfaiate/HySure}, dictionary learning-based approach, NSSR \cite{dong2016hyperspectral}\footnote{http://see.xidian.edu.cn/faculty/wsdong}, tensor-based approaches, STEREO \cite{kanatsoulis2018hyperspectral}\footnote{https://github.com/marhar19/HSR\_via\_tensor\_decomposition}, CSTF \cite{li2018fusing}\footnotemark, and LTTR \cite{dian2019learning}\footnotemark[\value{footnote}]\footnotetext{https://sites.google.com/view/renweidian}, and DL-based methods, unsupervised uSDN \cite{qu2018unsupervised}\footnote{https://github.com/aicip/uSDN} and supervised MHFnet \cite{xie2019multispectral}\footnote{https://github.com/XieQi2015/MHF-net}. As for the supervised deep method MHFnet, we use the remaining part of each dataset for the training following the strategies in \cite{xie2019multispectral}. 

\begin{table*}[!t]
    \centering
    \caption{The ability of learning unkonwn SRF and PSF of competing methods.}
    \resizebox{1\textwidth}{!}{ 
    \begin{tabular}{c|ccccccccc|cc|c}
        \toprule[1.5pt]
        Functions & ~GSA & CNMF & CSU & FUSE & HySure & NSSR & STEREO & CSTF & LTTR~ & ~uSDN & MHFnet & CUCaNet\\
        \hline\hline
        SRF & \xmark & \xmark & \xmark & \xmark & \cmark & \xmark & \xmark & \xmark & \xmark & \xmark & \cmark & \cmark \\
        PSF & \xmark & \xmark & \xmark & \xmark & \cmark & \xmark & \xmark & \xmark & \xmark & -      & \cmark & \cmark \\
        \bottomrule[1.5pt]
    \end{tabular}
    }
    \label{table:Dependency}
\end{table*}

Note that most of the above methods rely on the prior knowledge of SRFs and PSFs. We summarize the properties of all compared methods in learning SRFs and PSFs (see Table \ref{table:Dependency}), where only HySure and MHFnet are capable of learning the two unknown functions. More specifically, HySure adopts a multi-stage method and MHFnet models them as convolution layers under a supervised framework. Hence our CUCaNet serves as the first unsupervised method that can simultaneously learn SRFs and PSFs in an end-to-end fashion.

\begin{figure*}[!t]
    \centering
    \includegraphics[width=1\textwidth]{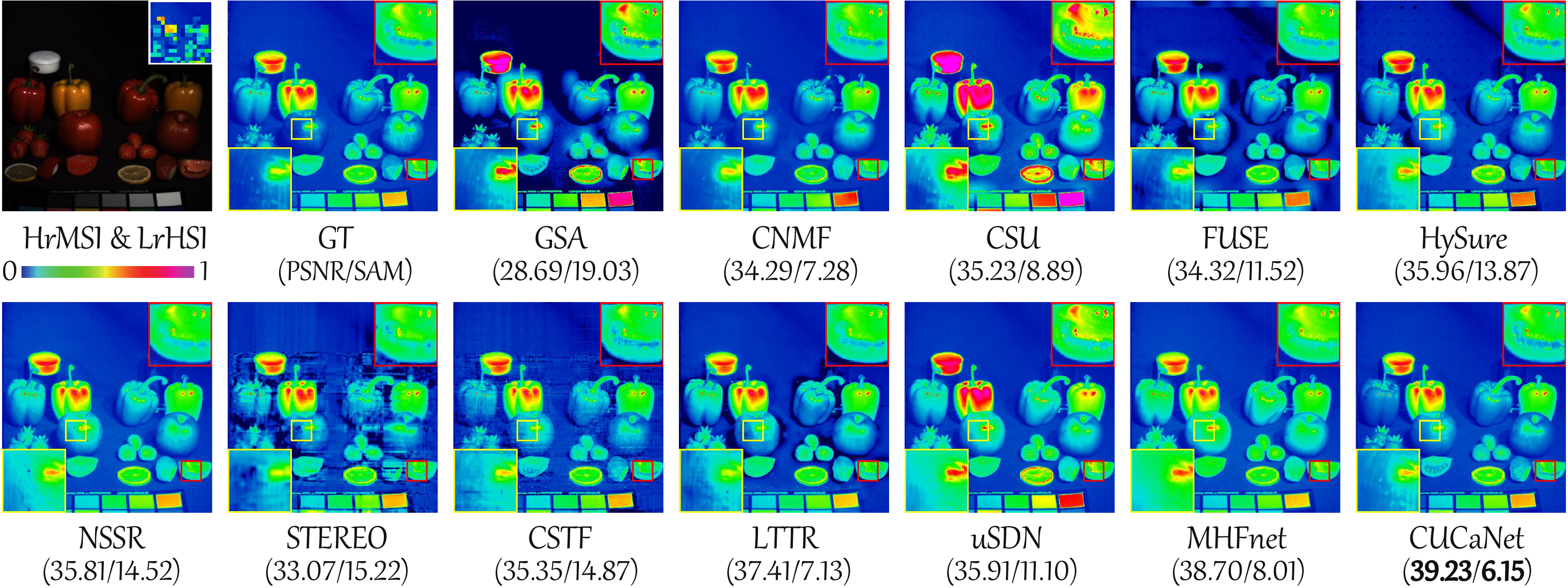}
    \caption{The HSI-SR performance on the CAVE dataset (\textit{fake and real food}) of CUCaNet in comparison with SOTA methods. For each HSI, the 20th (590nm) band image is displayed with two demarcated areas zoomed in 3 times for better visual assessment, and two main scores (PSNR/SAM) are reported with the best results in bold.}
    \label{fig:CAVE-1}
\end{figure*}

\begin{table*}[!t]
    \centering
    \caption{Quantitative performance comparison with the investigated methods on the CAVE dataset. The best results are shown in bold.}
    \resizebox{1\textwidth}{!}{
    \begin{tabular}{c|ccccccccc|cc|c}
        \toprule[1.5pt]
        \multirow{2}{*}{Metric} & \multicolumn{12}{c}{Method} \\
        \cline{2-13}
         & ~GSA~ & CNMF  & CSU   & ~FUSE & HySure& NSSR & STEREO& CSTF  & LTTR~ & ~uSDN & MHFnet & CUCaNet \\
        \hline\hline
        PSNR  & 27.89 & 30.11 & 30.26 & 29.87 & 31.26 & 33.52 & 30.88 & 32.74 & 35.45 & 34.67 & 37.30 & \textbf{37.51} \\
        SAM   & 19.71 & 9.98  & 11.03 & 16.05 & 14.59 & 12.09 & 15.87 & 13.13 & 9.69  & 10.02 & 7.75  & \textbf{7.49}  \\
        ERGAS & 1.11  & 0.69  & 0.65  & 0.77  & 0.72  & 0.69  & 0.75  & 0.64  & 0.53  & 0.52  & 0.49  & \textbf{0.47}  \\
        SSIM  & 0.713 & 0.919 & 0.911 & 0.876 & 0.905 & 0.912 & 0.896 & 0.914 & 0.949 & 0.921 & \textbf{0.961} & 0.959 \\
        UQI   & 0.757 & 0.911 & 0.898 & 0.860 & 0.891 & 0.904 & 0.873 & 0.902 & 0.942 & 0.905 & 0.949 & \textbf{0.955} \\
        \bottomrule[1.5pt]
    \end{tabular}
    }
    \label{table:CAVE}
\end{table*}

\textbf{Indoor Dataset.}~We first conduct experiments on indoor images of the CAVE dataset. The average quantitative results over 16 testing images are summarized in Table \ref{table:CAVE} with the best ones highlighted in bold. From the table, we can observe that LTTR and CSTF can obtain better reconstruction results than other conventional methods, mainly by virtue of their complex regularizations under tensorial framework. Note that the SAM values of earlier methods CNMF and CSU are still relatively lower because they consider the coupled unmixing mechanism. As for the DL-based methods, supervised MHFnet outperforms unsupervised uSDN evidently, while our proposed CUCaNet achieves the best results in terms of four major metrics. Only the SSIM value of ours is slightly worse than that of the most powerful competing method MHFnet, due to its extra exploitation of supervised information. 

\begin{figure*}[!t]
    \centering
    \includegraphics[width=1\textwidth]{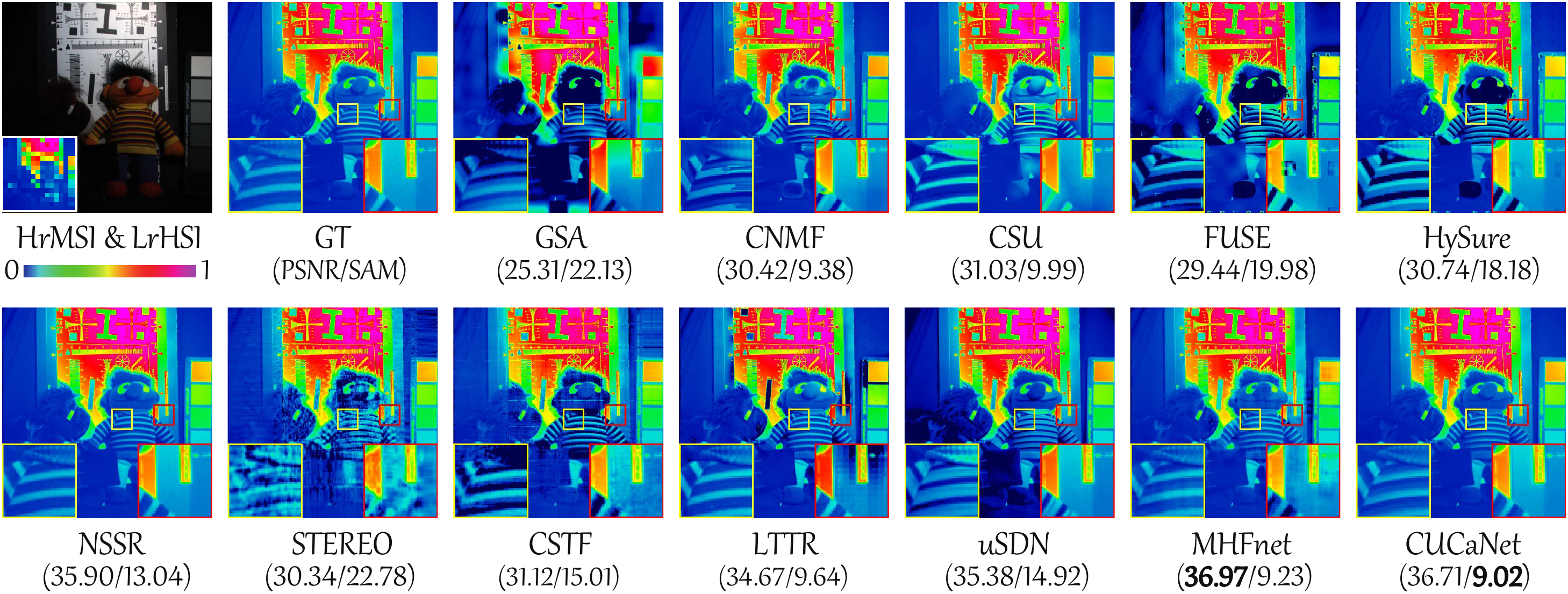}
    \caption{The HSI-SR performance on the CAVE dataset (\textit{chart and staffed toy}) of CUCaNet in comparison with SOTA methods. For each HSI, the 7th (460nm) band image is displayed with two demarcated areas zoomed in 3.5 times for better visual assessment, and two main scores (PSNR/SAM) are reported with the best results in bold.}
    \label{fig:CAVE-2}
\end{figure*}

The visual comparison on two selected scenes demonstrated in Fig. \ref{fig:CAVE-1} and Fig. \ref{fig:CAVE-2} exhibits a consistent tendency. From the figures, we can conclude that the results of CUCaNet maintain the highest fidelity to the groundtruth (GT) compared to other methods. For certain bands, our method can not only estimate background more accurately, but also maintain the texture details on different objects. The SAM values of CUCaNet on two images are obviously less than others, which validates the superiority in capturing the spectral characteristics via joint coupled unmixing and degrading functions learning.

\begin{figure*}[!t]
    \centering
    \includegraphics[width=1\textwidth]{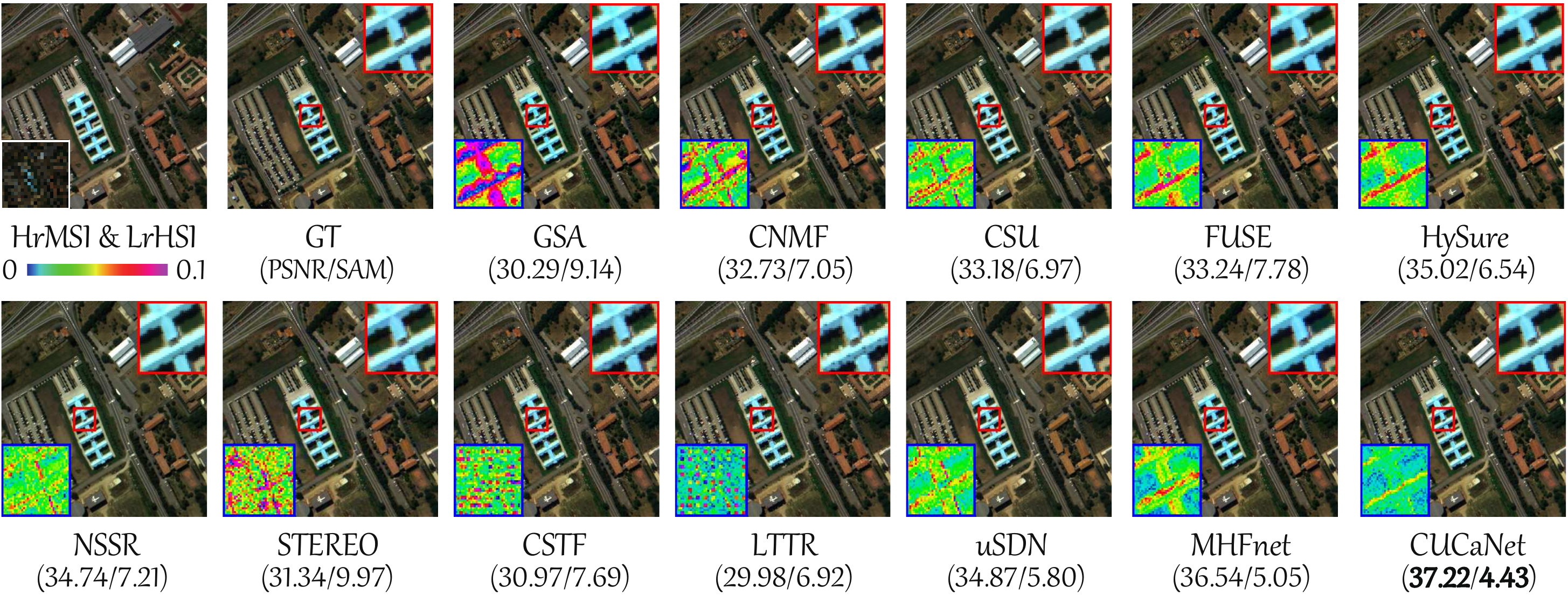}
    \caption{The HSI-SR performance on the Pavia University dataset (cropped area) of all competing methods. The false-color image with bands 61-36-10 as R-G-B channels is displayed. One demarcated area (red frame) as well as its RMSE-based residual image (blue frame) with respect to GT are zoomed in 3 times for better visual assessment.}
    \label{fig:paviau}
\end{figure*}

\textbf{Remotely Sensed Dataset.}~We then carry out more experiments using airborne HS data to further evaluate the generality of our method. The quantitative evaluation results on the Pavia University and Chikusei datasets are provided in Table \ref{table:paviau} and Table \ref{table:Chikusei}, respectively. Generally, we can observe a significant performance improvements than on CAVE, since more spectral information can be used as the number of HS bands increases. For the same reason, NMF-based and Bayesian-based methods show competitive performance owing to their accurate estimation of high-resolution subspace coefficients \cite{yokoya2017hyperspectral}. The limited performance of tensor-based methods suggests they may lack robustness to the spectral distortions in real cases. The multi-stage unsupervised training of uSDN makes it easily trapped into local minima, which results in only comparable performance to state-of-the-art conventional methods such as HySure and FUSE. It is particularly evident that MHFnet performs better on Chikusei rather than Pavia University. This can be explained by the fact that training data is relatively adequate on Chikusei so that the tested patterns are more likely to be well learned. We have to admit, however that MHFnet requires extremely rich training samples, which restricts its practical applicability to a great extent. Remarkably, our CUCaNet can achieve better performance in most cases, especially showing advantage in the spectral quality measured by SAM, which confirms that our method is good at capturing the spectral properties and hence attaining a better reconstruction of HrHSI.

\begin{table*}[!t]
    \centering
    \caption{Quantitative performance comparison with the investigated methods on the Pavia University dataset. The best results are shown in bold.}
    \resizebox{1\textwidth}{!}{ 
    \begin{tabular}{c|ccccccccc|cc|c}
        \toprule[1.5pt]
        \multirow{2}{*}{Metric} & \multicolumn{12}{c}{Method} \\
        \cline{2-13}
              & ~GSA~ & CNMF  & CSU   & ~FUSE & HySure& NSSR  & STEREO& CSTF  & LTTR~ & ~uSDN & MHFnet & CUCaNet \\
        \hline\hline
        PSNR  & 30.29 & 32.73 & 33.18 & 33.24 & 35.02 & 34.74 & 31.34 & 30.97 & 29.98 & 34.87 & 36.34 & \textbf{37.22} \\
        SAM   & 9.14  & 7.05  & 6.97  & 7.78  & 6.54  & 7.21  & 9.97  & 7.69  & 6.92  & 5.80  & 5.15  & \textbf{4.43}  \\
        ERGAS & 1.31  & 1.18  & 1.17  & 1.27  & 1.10  & 1.06  & 1.35  & 1.23  & 1.30  & 1.02  & 0.89  & \textbf{0.82}  \\
        SSIM  & 0.784 & 0.830 & 0.815 & 0.828 & 0.861 & 0.831 & 0.751 & 0.782 & 0.775 & 0.871 & \textbf{0.919} & 0.914 \\
        UQI   & 0.965 & 0.973 & 0.972 & 0.969 & 0.975 & 0.966 & 0.938 & 0.969 & 0.967 & 0.982 & 0.987 & \textbf{0.991} \\
        \bottomrule[1.5pt]
    \end{tabular}
    }
    \label{table:paviau}
\end{table*}

\begin{figure*}[!t]
    \centering
    \includegraphics[width=1\textwidth]{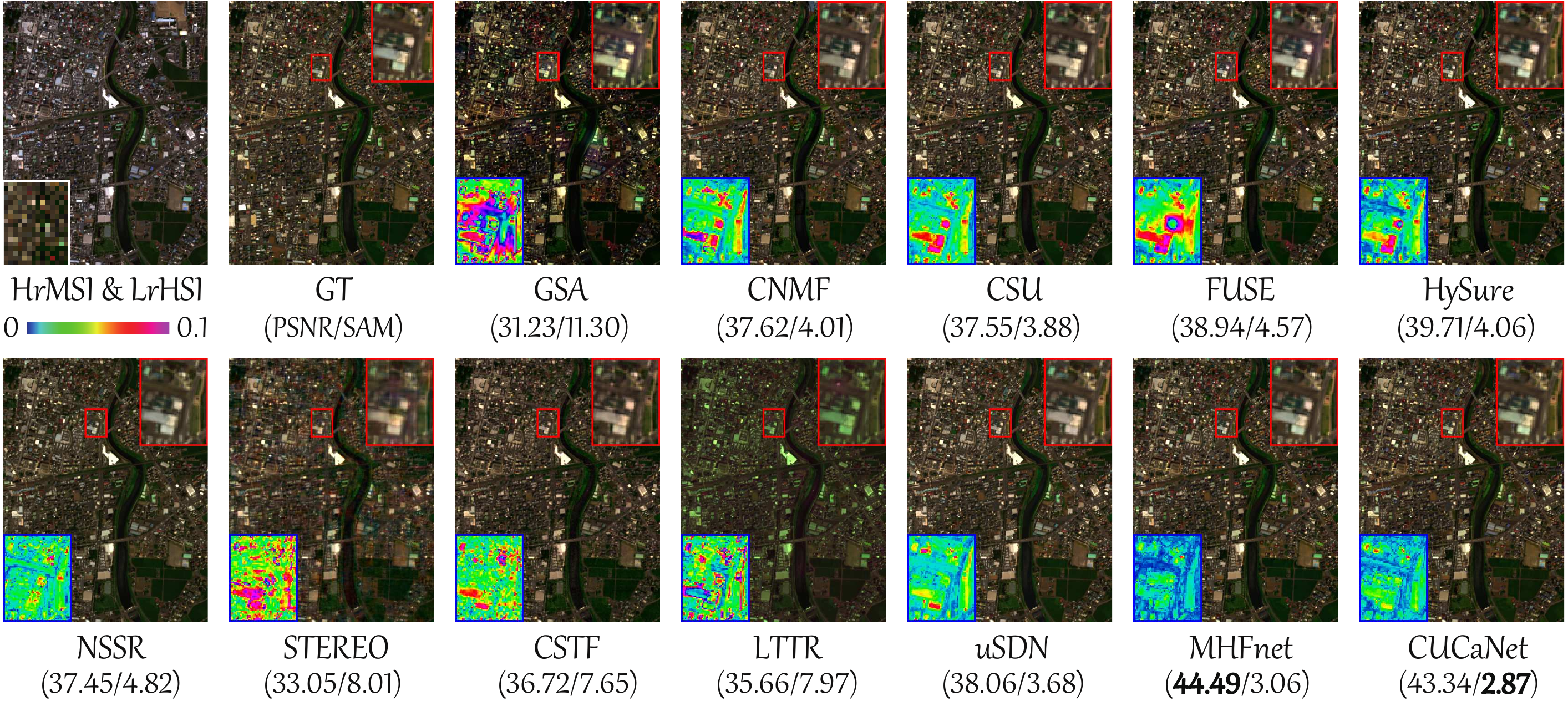}
    \caption{The HSI-SR performance on the Chikusei dataset (cropped area) of all competing methods. The false-color image with bands 61-36-10 as R-G-B channels is displayed. One demarcated area (red frame) as well as its RMSE-based residual image (blue frame) with respect to GT are zoomed in 3 times for better visual assessment.}
    \label{fig:Chikusei}
\end{figure*}

\begin{table*}[!t]
    \centering
    \caption{Quantitative performance comparison with the investigated methods on the Chikusei dataset. The best results are shown in bold.}
    \resizebox{1\textwidth}{!}{ 
    \begin{tabular}{c|ccccccccc|cc|c}
        \toprule[1.5pt]
        \multirow{2}{*}{Metric} & \multicolumn{12}{c}{Method} \\
        \cline{2-13}
              & ~GSA~ & CNMF  & CSU   & ~FUSE & HySure& NSSR  & STEREO& CSTF  & LTTR~ & ~uSDN & MHFnet & CUCaNet \\
        \hline\hline
        PSNR  & 32.07 & 38.03 & 37.89 & 39.25 & 39.97 & 38.35 & 32.40 & 36.52 & 35.54 & 38.32 & \textbf{43.71} & 42.70 \\
        SAM   & 10.44 & 4.81  & 5.03  & 4.50  & 4.35  & 4.97  & 8.52  & 6.33  & 7.31  & 3.89  & 3.51  & \textbf{3.13}  \\
        ERGAS & 0.98  & 0.58  & 0.61  & 0.47  & 0.45  & 0.63  & 0.74  & 0.66  & 0.70  & 0.51  & 0.42  & \textbf{0.40}  \\
        SSIM  & 0.903 & 0.961 & 0.945 & 0.970 & 0.974 & 0.961 & 0.897 & 0.929 & 0.918 & 0.964 & 0.985 & \textbf{0.988} \\
        UQI   & 0.909 & 0.976 & 0.977 & 0.977 & 0.976 & 0.914 & 0.902 & 0.915 & 0.917 & 0.976 & \textbf{0.992} & 0.990 \\
        \bottomrule[1.5pt]
    \end{tabular}
    }
    \label{table:Chikusei}
\end{table*}

Fig. \ref{fig:paviau} and Fig. \ref{fig:Chikusei} show the HSI-SR results demonstrated in false-color on these two datasets. Since it is hard to visually discern the differences of most fused results, we display the RMSE-based residual images of local windows compared with GT for better visual evaluation. For both datasets, we can observe that GSA and STEREO yield bad results with relatively higher errors. CNMF and CSU show evident patterns in residuals that are similar to the original image, which indicates that their results are missing actual details. The block pattern-like errors included in CSTF and LTTR make their reconstruction unsmooth. Note that residual images of CUCaNet and MHFnet exhibit more dark blue areas than other methods. This means that the errors are small and the fused results are more reliable. 

\section{Conclusion}
In this paper, we put forth CUCaNet for the task of HSI-SR by integrating the advantage of coupled spectral unmixing and deep learning techniques. For the first time, the learning of unknown SRFs and PSFs across MS-HS sensors is introduced into an unsupervised coupled unmixing network. Meanwhile, a cross-attention module and reasonable consistency enforcement are employed jointly to enrich feature extraction and guarantee a faithful production. Extensive experiments on both indoor and airborne HS datasets utilizing diverse simulations validate the superiority of proposed CUCaNet with evident performance improvements over competitive methods, both quantitatively and perceptually. Finally, we will investigate more theoretical insights on explaining the effectiveness of the proposed network in our future work.

\textbf{Acknowledgements.} 
This work has been supported in part by projects of the National Natural Science Foundation of China (No. 61721002, No. U1811461, and No. 11690011) and the China Scholarship Council.

\clearpage
%
%
\bibliographystyle{splncs04}
\bibliography{6628}
\end{document}